\documentclass[pra,aps,twocolumn,epsfig,nofootinbib]{revtex4}

\usepackage{amsmath}
\usepackage{amssymb}
\usepackage[dvips]{graphicx}
\usepackage{dcolumn}
\usepackage{bm}
\usepackage[colorlinks=false,dvipdfm]{hyperref} 
\usepackage{setspace}
\usepackage{amsfonts}
\usepackage{rotating}
\usepackage{makeidx}
\usepackage{amsthm}

\newcommand{\sgn}{\mbox{sgn}}

\begin{document}
	
	
	\title{Hear $\pi$ from Quantum Galperin Billiards\footnote{Canadian Journal of Physics 101, 491 (2023)}}

	\author{Yin Cai}
	\affiliation{Department of Physics, School of Science, Tianjin University, Tianjin 300072, China}
	
	\author{Fu-Lin Zhang}
	\email[Corresponding author: ]{flzhang@tju.edu.cn}
	\affiliation{Department of Physics, School of Science, Tianjin University, Tianjin 300072, China}

	\date{\today}

	\begin{abstract}
		Galperin introduced an interesting method to learn the digits of $\pi $  by counting the collisions of two billiard balls and a hard wall.	This paper studies two quantum versions of the Galperin billiards.	It is shown	that the digits of $\pi $ can be observed in the phase shifts of the quantum models.	\vspace{2ex}
		\\
		\noindent{\textbf{Key} \textbf{words:} Galperin billiards, phase shift, quantum classical correspondence, adiabatic approximation, scattering}
	\end{abstract}
\maketitle

\section{Introduction} \label{intro}

\begin{figure}[h]
\centering
\includegraphics[height=2cm]{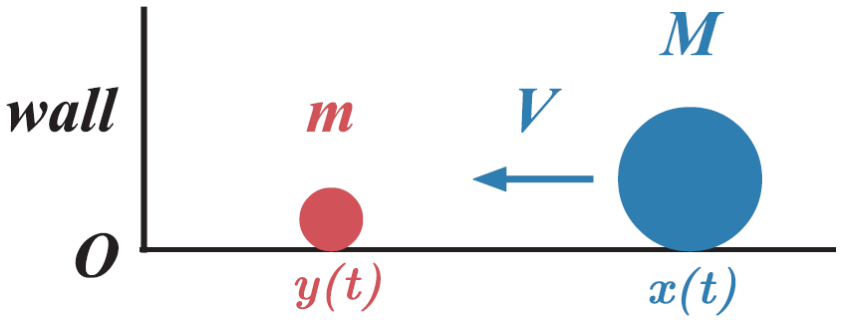}
\caption{Schematic picture of Galperin billiards, consisting of a big ball M, a small ball m and a wall.}
\label{figclassical}
\end{figure}

The number $\pi $ has attracted interest in human history.
There are various elegant ways to determine $\pi$ with a good precision;
for instance, the inscribed and circumscribed polygons on a circle \cite{1,2,3}
 and
the Monte Carlo Method based on computers.
Twenty years ago,
Galperin \cite{8,6} invented an extraordinary method to learn the digits of $\pi $, known as \emph{Galperin billiards}.
In such a scenario, as shown in Fig. \ref{figclassical}, a big ball is hurled towards a small ball that has its back to a wall.
An observer counts the ensuing elastic collisions until the big ball's momentum is redirected enough to fully turn around.
When the ratio of masses for the two billiard balls $\frac{M}{m}=100^{N}$,
 the total number of hits in the system is $[\pi10^{N}]$
 where $[\ ]$ is the greatest integer function.

The interesting results of Galperin billiards have been discussed in different aspects,
such as their relationships with geometrical optics and Grover algorithm for quantum search \cite{8,9}
 In this paper, we study two  quantum extensions of the Galperin billiards.
 In our first (\emph{semiclassical}) model, the motion of the small ball is assumed to be described by quantum mechanics,
 and  the big one is classical.
 In addition,  the classical part is assumed to move slowly, and hence the quantum part is treated under the adiabatic approximation.
In the second (\emph{fully quantum}) version, both balls are quantum, which is equivalent to a scattering problem
in a two-dimensional unbounded sector domain.
The phase shifts in the two models are shown to correspond to the number of collisions in classical mechanics.
As phase shifts in scattering problems are closely related with state densities and energy spectrums \cite{dai2009number,li2015heat},
this work connects the Galperin model with
the classic problem \emph{Can you hear the shape of a drum? } formulated by Mark Kac \cite{kac1966can}.
That is, the shape of two-dimensional unbounded sector domain,
corresponding to the quantum Galperin model,
determines the spectrum of a quantum particle in it and the digits of $\pi$ in collisions.

\section{Semiclassical model}\label{Semi}

\begin{figure}[tbph]
 \centering
\includegraphics[height=2cm]{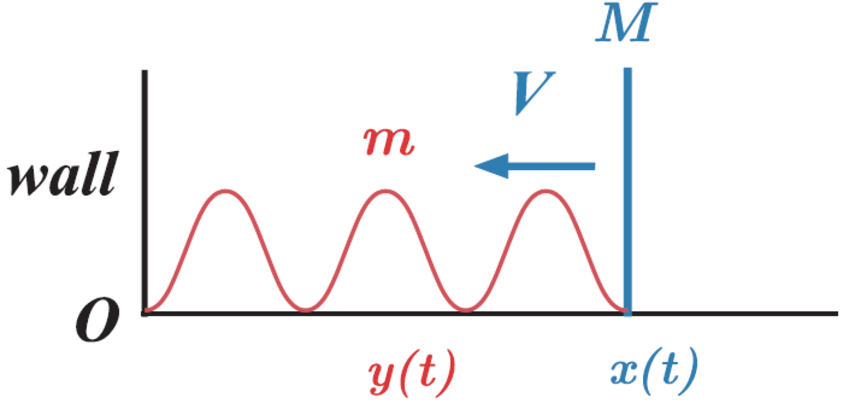}
\caption{Schematic picture of semiclassical Galperin billiards, consisting of the classical big ball M (blue vertical line on the right), the quantum small ball m, and a wall. The red curve between M and the wall shows the probability density for the wave function of m with $n$=3.}
\label{Well}
\end{figure}

We begin by replacing the small ball in the original model  with a quantum particle.
As shown in Fig. \ref{Well}, the classical  big ball and the wall constitute a one-dimensional infinite potential well, as the quantum small ball is situated between them.
The motions of the two balls are governed  by their interaction. Let $x$ and $y$ denote the positions of the big and small balls, respectively. The small ball satisfies the Schr\"odinger equation

\begin{flalign}
&i\hbar \frac{\partial }{\partial t}\Psi \left( y,t\right) =\left[ -\frac{%
\hbar ^{2}}{2m}\frac{\partial ^{2}}{\partial y^{2}}+V\left( y\right) \right]
\Psi \left( y,t\right),& \label{0.0}
\end{flalign}
where $m$ is its mass and

\begin{flalign}
&V(y)=\left\{
\begin{array}{c}
0,\ \ \ \ 0\leqslant y\leqslant x, \\
+\infty ,\ \mathrm{otherwise}.%
\end{array}%
\right.&   \label{0.1}
\end{flalign}


%

Here, we assume that the big ball moves very gradually, and therefore the wave function of the quantum particle changes adiabatically.
Under the adiabatic approximation, we assume that the position coordinate $x$
of the big ball is fixed, and the wave function of the small ball is given by
\begin{flalign}
&\Psi =\sum_{n=1}^{+ \infty} C_{n} \Psi _{n} e^{-i\frac{1}{\hbar }%
\int_{0}^{t}E_{n} dt+i\gamma _{n}},&  \label{1.1}
\end{flalign}%
where $E_n= \frac{n^{2}\pi ^{2}\hbar ^{2}}{2 m x^{2}}$,$\Psi _{n}=\sqrt{\frac{2}{x}}  \sin \frac{n\pi y }{x}$, and $\gamma_{n} $ is the Berry phase gained by the eigenstate $\Psi _{n}$ as the big ball coordinate has changed from the initial position to the position $x$. Under the adiabatic approximation, as the Hamiltonian gradually changes, the wave function will get an additional phase factor, and so called the Berry phase \cite{12}.
 Here, both the initial and final values of $x$ (big ball position) are $+ \infty$, and the final time $t \rightarrow + \infty$. The kinetic energy of the quantum small ball is close to zero at the beginning and end.
 The Berry phase for the $n$th eigenstate can be
 \begin{flalign}
&\gamma _{n}=i\int_{0}^{t}{\langle }\Psi _{n}|\frac{d}{dt}|\Psi _{n}{\rangle }%
d t =0,&\label{1.6}
\end{flalign}
 as ${\langle }\Psi _{n}| d |\Psi _{n}{\rangle } =0$
(see Supplementary material
for more details).

To observe the motion of the quantum particle,
we consider a superposition of two adjacent levels as
\begin{flalign}
&\Psi =\frac{1}{\sqrt{2}}(\Psi _{n} e^{-i\frac{1}{\hbar }%
\int_{0}^{t}E_{n} dt } +\Psi _{n+1}  e^{-i\frac{1}{\hbar }%
\int_{0}^{t}E_{n+1} dt }  ).&  \label{2.1}
\end{flalign}
Its average position is given by (see Supplementary material
for more details)
\begin{small}
\begin{flalign}
\overline{ y } &={\langle }\Psi |y|\Psi {\rangle }  \notag \\
&=\frac{x}{2}-\frac{x}{\pi ^{2}}\frac{8n(n+1)}{(2n+1)^{2}}\cos \int_{0}^{t}%
\frac{E_{n+1}-E_{n}}{\hbar }dt. & \label{2.3}
\end{flalign}
\end{small}
The average position
 vibrates around the midpoint $\frac{x}{2}$ at the Bohr frequency $\frac{E_{n+1}-E_{n}}{\hbar }$.
Hence, the phase difference, divided by $ \pi$,  between the two eigenenstates can be regarded as the quantum correspondence of the number of collisions in classical mechanics.

The energy of the quantum ball
is
\begin{flalign}
&E=\frac{1}{2}(E_{n}+E_{n+1})
=\frac{2n^{2}+2n+1}{2}\frac{\pi ^{2}\hbar ^{2}}{%
m x^{2}}.&  \label{2.2}
\end{flalign}
We assume that the speed change of the classical ball is determined by its energy exchange with the quantum particle.
One can obtain
 \begin{flalign}
&\frac{2n^{2}+2n+1}{2}\frac{\pi ^{2}\hbar ^{2}}{m  }(\frac{1}{x_{\min }^{2}}-\frac{1%
}{x^{2}})=\frac{1}{2}M v^{2},& \label{2.7}
\end{flalign}
where $v=\frac{dx}{dt }$, $M$ is the mass of the classical ball, and $x_{\min }$ is its retracing point.
Then, the definite integral in (\ref{2.3}) can be solved by  variable transformation from $t$ to $x$.
It is obtained as
\begin{flalign}
&\int_{0}^{+ \infty }\frac{E_{n+1}-E_{n}}{\hbar }dt=\sqrt{\frac{4n^{2}+4n+1}{%
4n^{2}+4n+2}}\pi ^{2}R,&  \label{2.9}
\end{flalign}
where $R=\sqrt{\frac{M}{m}}$ is the square root of the two masses ratios.
The number of the cosine function in  (\ref{2.3}) reaching its extreme values is $\sqrt{\frac{4n^{2}+4n+1}{4n^{2}+4n+2}}\pi R$,
which  does not depend on $x_{\min }$, or the initial speed of the classical ball.

\begin{figure}[tbph]
\centering
\includegraphics[width=9cm]{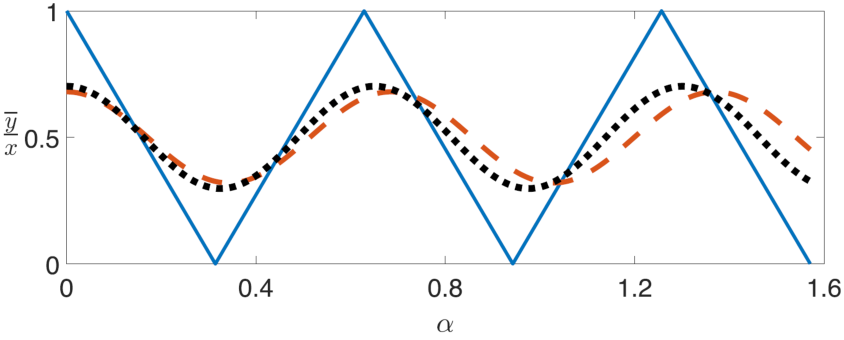}
\caption{
A comparison between the average position $\bar{y}$ and the corresponding results of a classical model when $\beta=\pi/10$.
The three curves are obtained for the classical model (solid blue line),
the semiclassical model with quantum number  $n=1$ (dashed red line),
and
$n=10$ (dotted black line).
}
\label{SYX}
\end{figure}

In the classical model, the full change of phase of  \emph{action-angle}  variables is $\pi^{2}R$ \cite{11} and the collision times is $[\frac{\pi}{\beta}]$ with $ \beta = \mathrm{arccot} R$ \cite{8}.
When $n \rightarrow + \infty$ and $R \rightarrow + \infty$,
the phase difference (\ref{2.9}) in the semiclassical model is equal to the full change of phase in the classical model. At the same time, the number of extremums in the semiclassical model is equal to the collision times in the classical model.
We define a \emph{normalized} time as
\begin{flalign}
  &\alpha =\sgn(v)  \arccos \frac{\rho_{\min }}{\rho},&
\label{2.5}
\end{flalign}
where sgn is the sign function, $\rho =\sqrt{M}x$ in the semiclassical model while $\rho =\sqrt{Mx^{2}+my^{2}}$ in the classical one, and the values of $\rho_{\min }$ are their minimums. Here, we mix $y$ in $\rho $ for the classical model to avoid the case that $x$ may remain stationary in a finite time interval at its retracing point. When $M\gg m$, the two definitions of $\rho $ consistent with each other.
As $t$ increases  from $0$ to  $+ \infty$,
the value of  $\alpha $  changes continuously from $ -\frac{\pi }{2}$ to  $ \frac{\pi }{2}$,
 and $\rho$   decreases from $+ \infty$ to $\rho_{\min }$ and then  goes back to $+ \infty$.
Using $  \alpha $ as the variable, in Fig. \ref{SYX},  we
show $x$ comparison between the average position $\bar{y}$ and the result of a corresponding classical model.
In general, the oscillation of $\bar{y}$ correspond to the collisions in classical mechanics.
The period of the quantum  oscillation is greater than the one of the classical collisions,
 while the difference between them is reduced as the quantum number $n$  increases.

 \section{Fully Quantum Model}

We now consider a fully quantum model in which both balls are described by quantum mechanics.
The stationary Schr\"odinger equation of the fully quantum model can be written as
\begin{flalign}
&\left[ -\frac{\hbar ^{2}}{2M}\frac{\partial ^{2}}{\partial x^{2}}-\frac{%
\hbar ^{2}}{2m}\frac{\partial ^{2}}{\partial y^{2}}+V(x,y)\right] \Psi
(x,y)=E\Psi (x,y).&  \label{3.0}
\end{flalign}
where $x$ and $y$ denote respectively the positions of the big ball of mass $M$ and the small ball of mass $m$, respectively and
\begin{flalign}
&V(x,y)=\left\{
\begin{array}{c}
0,\ \  \ \ 0\leqslant y \leqslant x, \\
+\infty ,\  \mathrm{otherwise}.
\end{array}%
\right.&
\end{flalign}%
Such a two-body problem in one-dimensional space can be  converted into a free particle in a two-dimensional  unbounded sector domain.
Namely,  introducing the
polar coordinates
\begin{flalign}
&\sqrt{M}x =\rho \cos \theta  , \  \
&\sqrt{m}y =\rho \sin \theta ,&\label{3.1}
\end{flalign}%
and the Schr\"odinger equation turns to  (see Supplementary
material for more details)
\begin{flalign}
 &\! \left[-\frac{\hbar ^{2}}{2 } \! \left(\! \frac{\partial ^{2}}{\partial \rho^{2}} \!+ \!
\frac{1}{\rho }\frac{\partial }{\partial \rho } \!+\! \frac{1}{\rho ^{2}}\frac{
\partial ^{2}}{\partial \theta ^{2}} \! \right)\! +\! V( \theta  )\! \right]\Psi (\rho ,\theta )=E\Psi (\rho
,\theta ),& \label{3.3}
\end{flalign}%
where
\begin{flalign}
&V(\theta )=\left\{
\begin{array}{l}
0,\ \ \ \ 0\leqslant \theta \leqslant \beta=\arctan \sqrt{\frac{m}{M}}, \\
+\infty ,\ \mathrm{otherwise}%
\end{array}%
\right. &  \label{3.4}
\end{flalign}
When $M\gg m$,
$\rho \approx \sqrt{M} x$ and $\theta/\beta \approx  {y}/{ x}$.
The former approaches to the position of the big ball, and the latter becomes the (normalized) position of the small one corresponding to the curves in Fig. \ref{SYX}.

\begin{figure}
\centering
\includegraphics[height=2.2cm]{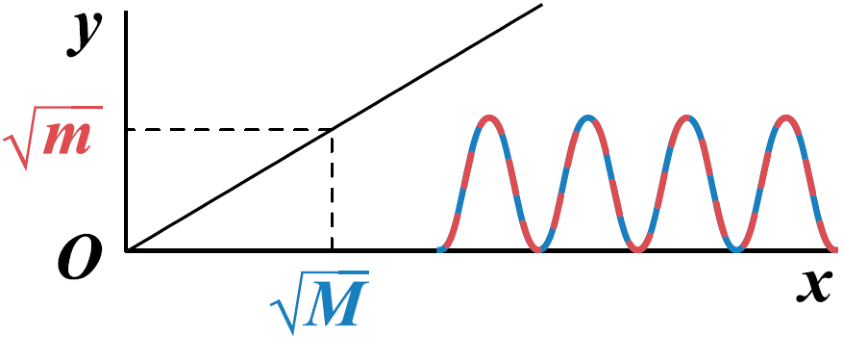}
\caption{ The fully quantum Galperin model is equivalent to a free particle in a two-dimensional unbounded sector domain in the first quadrant where $\frac{y}{x}\in[0,\sqrt{\frac{m}{M}}]$.
}
\label{fig_sim}
\end{figure}

As the potential $V(\theta)$ does not depend on the radial coordinate $\rho$,
the stationary Schr\"odinger eq.(\ref{3.3}) can be solved by using the separation of variables method.
We assume that
$\Psi (\rho ,\theta )=R(\rho )\Phi (\theta )$ and  $\Phi (\theta )=0$ when $\theta \notin [0,\beta]$.
The Schr\"odinger equation turns to

\begin{flalign}
&-  \frac{\partial ^{2}}{\partial
\theta ^{2}}\Phi (\theta ) = l^2 \Phi (\theta ),
\label{3.7a} \\
 &-\frac{\hbar ^{2}}{2  } \left(\frac{\partial ^{2}}{\partial \rho ^{2}}+\frac{1}{%
\rho }\frac{\partial }{\partial \rho }- \frac{l^2}{\rho ^{2}}  \right)R(\rho ) =E R(\rho).&
\label{3.7b}
\end{flalign}
The eigenfunctions and  eigenvalues for the angular part can be directly found to be
$ \Phi (\theta ) = \sin l\theta  $ when $\theta \in [0,\beta]$ and  $l=  {n\pi  }/{\beta }$ with $n=1,2,3...$
The physical solution to the radial eq.(\ref{3.7b}) is a linear combination of the first kind of Bessel function, as the second kind of Bessel function (Neumann function) tends to $- \infty$ when $\rho \rightarrow 0$ \cite{7}.
Here, we write the common eigenfunctions in terms of the Hankel functions as
\begin{flalign}
&\Psi_{k,l}(\rho,\theta)=%
[H_{l}^{(1)}(k\rho)+H_{l}^{(2)}%
(k\rho)]\sin l\theta,&\label{3.27}%
\end{flalign}
with $k= {\sqrt{2  E }}/{\hbar }$  (see Supplementary
material for more details).

The probability current density of the stationary wave function (\ref{3.27})
can be divided into two equal and opposite parts without cross terms  (see Supplementary
material for more details),
which are along the radial direction and correspond to the two terms $H_{l}^{(1)}(k\rho) \sin l\theta $ and  $H_{l}^{(2)}(k\rho) \sin l\theta $ respectively.
In addition, when $\rho
\rightarrow\infty$ \cite{7},
\begin{flalign}
H_{l}^{(1)}(k\rho) &  \approx \sqrt{\frac{2}{\pi k\rho}}%
e^{i[(k\rho-(l +\frac{1}{2})\frac{\pi}{2}]},\label{3.30a}\\
H_{l}^{(2)}(k\rho) &  \approx \sqrt{\frac{2}{\pi k\rho}}%
e^{-i [k\rho-(l+\frac{1}{2})\frac{\pi}{2}]},\label{3.30b}&
\end{flalign}
which become spherical waves in the  two-dimensional sector domain.
Therefore, the two terms, $H_{l}^{(1)}(k\rho) \sin l\theta $ and  $H_{l}^{(2)}(k\rho) \sin l\theta $
 are incident and outgoing waves, respectively.
 They can be regarded as the quantum correspondences of
 the two movement processes in the classical model divided by  the retracing point of the big ball.
 Here, an observation is that, although a more reasonable incidence wave should promulgate along $x$-axis,
 the incident direction makes a difference of  at most $1$
  on
 the number of  classical reflections in the unbounded sector domain in Fig. \ref{fig_sim} \cite{8}.
%
The phase shift between the incident and outgoing waves
 can be directly derived from the asymptotic formulations (\ref{3.30a}) and (\ref{3.30b}) as
\begin{flalign}
&\delta(n) =(l+\frac{1}{2})\pi   =(\frac{n\pi }{\beta } + \frac{1}{2} )\pi. & \label{6.29a}
\end{flalign}
It does not depend on the wavenumber $k$, or energy $E$ equivalently,
and
can be regarded as the phase gained by the standing wave $\sin l \theta$.
The difference between  two adjacent phase shifts is
\begin{flalign}
&\Delta \delta=  \delta(n+1)-\delta(n) =\frac{ \pi^2 }{\beta }.&
\end{flalign}
When $M \gg m$, it becomes $\Delta \delta =   \pi^2 R$,
which  corresponds to the phase difference (\ref{2.9}) in the semiclassical model and the change of phase in the classical model.

To observe the reciprocal motion of $y$ between $0$ and $x$, i.e., $\theta$ between $0$ and $\beta$,
we consider a superposition of two eigenfunctions with the same energy and adjacent angular quantum number $l$ as
\begin{flalign}
&\Psi= \Psi_{k,l}+e^{i c\pi}\Psi_{k,l'},&
\end{flalign}
where $l={n\pi }/{\beta }$ ,$l'={(n+1)\pi} /{\beta }$ and $c=\frac{\pi}{2\beta }$.
Here, the relative phase $e^{i c\pi}$ ensures the phase consistency between the average position in quantum model and classsical position.
We also divide it into incident and outgoing parts according to the superscripts of Hankel functions.
The incident wave reads
\begin{flalign}
&\Psi ^{\mathrm{in}}= H_{l}^{(1)}(k\rho)\sin l \theta  +e^{i c\pi} H_{l'}^{(1)}(k\rho)\sin l'\theta, & \label{3.33}
\end{flalign}
and the outgoing term has the same form with the Hankel functions of the first kind replaced with the second kind.
For a fixed $\rho$,
the average values of $\theta $ in the incident wave is
\begin{small}
\begin{flalign}
&\overline{\theta }(\rho )=\frac{\int {\Psi ^{\mathrm{in}}}^{\ast }\theta
\Psi ^{\mathrm{in}}d\theta }{\int {\Psi ^{\mathrm{in}}}^{\ast }\Psi ^{%
\mathrm{in}}d\theta }  \nonumber \\
&=\frac{\beta }{2}-\frac{\beta }{\pi ^{2}}\frac{8n(n+1)}{(2n+1)^{2}}%
\frac{e^{i c\pi}H_{l}^{\left( 2\right) }H_{l^{\prime }}^{\left( 1\right)
}+e^{-i c\pi}H_{l}^{\left( 1\right) }H_{l^{\prime }}^{\left( 2\right) }}{{%
H_{l}^{\left( 1\right) }H_{l}^{\left( 2\right) }+H_{l^{\prime }}^{\left(
1\right) }H_{l^{\prime }}^{\left( 2\right) }}},&\ \ \   \label{3.35}
\end{flalign}%
\end{small}
where we omit the argument $k\rho$ to the Hankel functions and the result in outgoing wave for simplicity.
The expression (\ref{3.35}) is similar to the average position (\ref{2.3}) of the quantum particle in the semiclassical model.
In addtion,
the  vibration of the average value $\overline{\theta}(\rho) $
with $\rho$
 depends  on the relative phase between $H_{l}^{(1)}$ and $H_{l' }^{(2)}$,
 since $H_{\gamma}^{(1)*}=H_{\gamma}^{(2)}$ and $|H_{\gamma}^{(1)} (k\rho)|$  is a monotonous decrease function of $k\rho$.

\begin{figure}
\centering
\includegraphics[width=9cm]{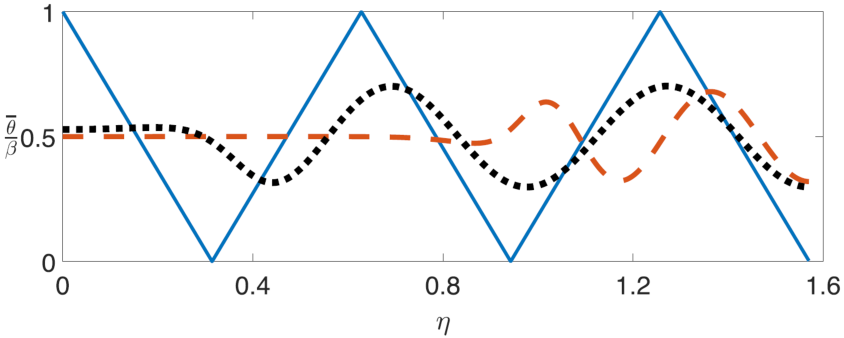}
\caption{A comparison between the average position $\bar{\theta}$ and the corresponding results of a classical model when $\beta=\pi/10$.
The three curves are obtained for the classical model (solid blue line)
and the fully quantum model with quantum number  $l=\frac{\pi}{\beta}$ (dashed red line) and  $l=\frac{10\pi}{\beta}$ (dotted black line).
 }
\label{figquantumline}
\end{figure}

In Fig. \ref{figquantumline}, we show a comparison between the average value  $\overline{\theta}(\rho) $
and the classical motion of $\theta$, as functions of
\begin{flalign}\label{eta}
&\eta =\arccos \frac{l}{k\rho }.&
\end{flalign}
As $\eta$ increases from $0$ to  $\frac{\pi}{2}$,
the value of  $\rho$  changes continuously from $ \frac{l}{k }$ to  $ + \infty$.
Here, $\frac{l}{k }$ is the quantum correspondence of the minimum of $\rho$ in the classical model,
i.e., the retracing point of the classical model $\rho= \rho_{\min }$.
Thus, the parameter  $\eta$ in the classical model is defined by replacing  $\frac{l}{k }$ with $\rho_{\min }$ in eq.(\ref{eta}).
%
Although the value of $\rho$ is allowed to be less than $\frac{l}{k }$ in the fully quantum model,
there is  no vibration of $\overline{\theta}(\rho)$ in the region of  $\rho< \frac{l}{k }$.
There is a obvious stationary region of the fully quantum model where $\eta$ is near zero.
The region becomes smaller as the quantum number $l$ increases. When $l\rightarrow + \infty$, this region almost disappears, and the period of the quantum oscillation is consistent with the one of the classical collisions.

 \section{Summary}\label{summ}

 In summary, we study two quantum extensions of the Galperin billiards, one of which  is semiclassical and the other is fully quantum.
In addition, we assume the small quantum ball in the semiclassical model, can be treated under the adiabatic approximation as the  big classical ball moves adiabatically.
The phase shifts in the two models are shown to correspond to the number of collisions in classical mechanics.

As the quantum model  connects the
digits of $\pi$ in the classical Galperin model  with
the spectrum of a quantum particle in a two-dimensional unbounded sector domain,
it would be intriguing to
study the relationship among  the digits of $\pi$ and  other physical problems in the same domain,
such as statistical mechanics and phonon eigenfunctions.
Actually,  these physical problems in similar domains are attracting attention \cite{zhang2020statistical,zhang2021phonon} .

\setlength{\parindent}{0pt}
\section*{Acknowledgements}
This work was supported by the National Natural Science Foundation of China (Grant Nos. 11675119, 11575125, and 11105097).
%
%
%
%
%
%
%
%
%
%
%
%
%
%
%
%
%
%

\bibliographystyle{unsrt}
\bibliography{QGalperin}

\begin{appendix}

\begin{widetext}



\textbf{Supplementary material}

	\section*{Proof of the relation ${\langle }\Psi _{n}| d |\Psi _{n}{\rangle } =0$}
	
	The eigenstate $\Psi _{n}$ is%
	\begin{equation}
		\Psi _{n}\left( y\right)=\sqrt{\frac{2}{x}}\sin \frac{n\pi y}{x}.  \tag{S1}
	\end{equation}%
	The derivative with respect to $t$, we obtain%
	\begin{align}
		d\Psi _{n}& =\sin \frac{n\pi y}{x}\cdot (-\frac{1}{\sqrt{2x^{3}}})dx+\sqrt{%
			\frac{2}{x}}(-\frac{n\pi y}{x^{2}})\cos \frac{n\pi y}{x}\cdot dx  \notag \\
		& =[-\frac{1}{\sqrt{2x^{3}}}\sin \frac{n\pi y}{x}-\frac{\sqrt{2}n\pi y}{%
			\sqrt{x^{5}}}\cos \frac{n\pi y}{x}]dx  \notag \\
		& =-\frac{1}{\sqrt{2x^{3}}}[\sin \frac{n\pi y}{x}+\frac{2n\pi y}{x}\cos
		\frac{n\pi y}{x}]dx.  \tag{S2}
	\end{align}%
	Therefore%
	\begin{equation}
		{\langle }\Psi _{n}|d|\Psi _{n}{\rangle =}-\frac{1}{x^{2}}%
		dx\int_{0}^{x}[\sin ^{2}\frac{n\pi y}{x}+\frac{n\pi y}{x}\sin \frac{2n\pi y}{%
			x}]dy.  \tag{S3}  \label{a3}
	\end{equation}
	The first integral of the above formula is%
	\begin{align}
		\int_{0}^{x}\sin ^{2}\frac{n\pi y}{x}dy &=\int_{0}^{x}\frac{1}{2}\left(
		1-\cos \frac{2n\pi y}{x}\right) dy  \notag \\
		&=\frac{x}{2}-\frac{1}{2}\int_{0}^{x}\cos \frac{2n\pi y}{x}dy  \notag \\
		&=\frac{x}{2},  \tag{S4}  \label{a4}
	\end{align}%
	the second is%
	\begin{align}
		&\int_{0}^{x}\frac{n\pi y}{x}\sin \frac{2n\pi y}{x}dy  \notag \\
		&=-\frac{x}{2n\pi }\int_{0}^{x}\frac{n\pi y}{x}d\cos \frac{2n\pi y}{x}
		\notag \\
		&=-\frac{x}{2n\pi }\left[ \left( \cos \frac{2n\pi y}{x}\frac{n\pi y}{x}%
		\right) |_{0}^{x}-\frac{n\pi }{x}\int_{0}^{x}\cos \frac{2n\pi y}{x}dy\right]
		\notag \\
		&=-\frac{x}{2}.  \tag{S5}  \label{a5}
	\end{align}%
	Substitution Eq.(\ref{a4}) and Eq.(\ref{a5}) into Eq.(\ref{a3}) gives%
	\begin{align}
		{\langle }\Psi _{n}|d|\Psi _{n}{\rangle }& =-\frac{1}{x^{2}}%
		dx\int_{0}^{x}[\sin ^{2}\frac{n\pi y}{x}+\frac{n\pi y}{x}\sin \frac{2n\pi y}{%
			x}]dy  \notag \\
		& =-\frac{1}{x^{2}}dx\left( \frac{x}{2}-\frac{x}{2}\right) =0.  \tag{S6}
	\end{align}
	
	\section*{The average position of the small-ball}
	
	We consider a superposition of two adjacent levels as%
	\begin{align}
		\Psi &=\frac{1}{\sqrt{2}}(\Psi _{n}+\Psi _{n+1})  \tag{S7} \label{a7} \\
		&=\sqrt{\frac{1}{x}}(\sin \frac{n\pi y}{x}\cdot e^{-i\int_{0}^{t}\frac{E_{n}%
			}{\hbar }dt}+\sin \frac{(n+1)\pi y}{x}\cdot e^{-i\int_{0}^{t}\frac{E_{n+1}}{%
				\hbar }dt}).  \notag
	\end{align}%
	Then let us calculate the average position of the small ball.
	\begin{align}
		\overline{|y|}& ={\langle }\Psi |y|\Psi {\rangle }  \tag{S8} \label{a8}\\
		& =\int_{0}^{x}[\sqrt{\frac{1}{x}}(\sin \frac{n\pi y}{x}\cdot
		e^{i\int_{0}^{t}\frac{E_{n}}{\hbar }dt}+\sin \frac{(n+1)\pi y}{x}\cdot
		e^{i\int_{0}^{t}\frac{E_{n+1}}{\hbar }dt})]y  \notag \\
		& \cdot \lbrack \sqrt{\frac{1}{x}}(\sin \frac{n\pi y}{x}\cdot
		e^{-i\int_{0}^{t}\frac{E_{n}}{\hbar }dt}+\sin \frac{(n+1)\pi y}{x}\cdot
		e^{-i\int_{0}^{t}\frac{E_{n+1}}{\hbar }dt})]dy  \notag \\
		& =\frac{1}{x}\int_{0}^{x}\sin ^{2}\frac{n\pi y}{x}ydy+\frac{1}{x}%
		\int_{0}^{x}\sin ^{2}\frac{(n+1)\pi y}{x}ydy  \notag \\
		& +\frac{1}{x}\int_{0}^{x}[\sin \frac{n\pi y}{x}\sin \frac{(n+1)\pi y}{x}%
		y]dy\cos \int_{0}^{t}\frac{E_{n}-E_{n+1}}{\hbar }dt.  \notag
	\end{align}%
	The first integral of the above formula is%
	\begin{small}
		\begin{align}
			&\frac{1}{x}\int_{0}^{x}\sin ^{2}\frac{n\pi y}{x}ydy  \tag{S9} \\
			&=\frac{1}{2x}\int_{0}^{x}\left( 1-\cos \frac{2n\pi y}{x}\right) ydy  \notag
			\\
			&=\frac{1}{2x}\left( \int_{0}^{x}ydy-\int_{0}^{x}\cos \frac{2n\pi y}{x}%
			ydy\right)  \notag \\
			&=\frac{1}{2x}\left( \frac{x^{2}}{2}-\frac{x}{2n\pi }\int_{0}^{x}yd\sin
			\frac{2n\pi y}{x}\right)  \notag \\
			&=\frac{1}{2x}\left( \frac{x^{2}}{2}-\frac{x}{2n\pi }\left[ \left( y\sin
			\frac{2n\pi y}{x}\right) |_{0}^{x}-\int_{0}^{x}\sin \frac{2n\pi y}{x}dy%
			\right] \right)  \notag \\
			&=\frac{x}{4}.  \notag
		\end{align}%
	\end{small}
	The second one is%
	\begin{small}
		\begin{align}
			&\frac{1}{x}\int_{0}^{x}\sin ^{2}\frac{(n+1)\pi y}{x}ydy  \tag{S10} \\
			&=\frac{1}{2x}\int_{0}^{x}\left( 1-\cos \frac{2(n+1)\pi y}{x}\right) ydy
			\notag \\
			&=\frac{1}{2x}\left( \int_{0}^{x}ydy-\int_{0}^{x}\cos \frac{2(n+1)\pi y}{x}%
			ydy\right)  \notag \\
			&=\frac{1}{2x}\left( \frac{x^{2}}{2}-\frac{x}{2n\pi }\int_{0}^{x}yd\sin
			\frac{2(n+1)\pi y}{x}\right)  \notag \\
			&=\frac{1}{2x}\left( \frac{x^{2}}{2}-\frac{x}{2n\pi }\left[ \left( y\sin
			\frac{2(n+1)\pi y}{x}\right) |_{0}^{x}-\int_{0}^{x}\sin \frac{2(n+1)\pi y}{x}%
			dy\right] \right)  \notag \\
			&=\frac{x}{4}.  \notag
		\end{align}%
	\end{small}
	The integral in the third term of Eq.(\ref{a7}) is%
	\begin{small}
		\begin{align}
			&\int_{0}^{x}y\sin \frac{n\pi y}{x}\sin \frac{(n+1)\pi y}{x}dy  \tag{S11}
			\\
			&=\int_{0}^{x}y\sin \frac{n\pi y}{x}\sin \frac{(n+1)\pi y}{x}dy  \notag \\
			&=-\frac{1}{2}\int_{0}^{x}\left[ y\cos \frac{\left( 2n+1\right) \pi y}{x}%
			-y\cos \frac{\pi y}{x}\right] dy  \notag \\
			&=-\frac{1}{2}\left[ \frac{x}{\left( 2n+1\right) \pi }\int_{0}^{x}yd\sin
			\frac{\left( 2n+1\right) \pi y}{x}-\frac{x}{\pi }\int_{0}^{x}yd\sin \frac{%
				\pi y}{x}\right]  \notag \\
			&=-\frac{1}{2}\left[ \frac{x}{\left( 2n+1\right) \pi }\left[ \frac{-2x}{%
				\left( 2n+1\right) \pi }\right] -\frac{x}{\pi }\left( \frac{-2x}{\pi }%
			\right) \right]  \notag \\
			&=x^{2}\left[ \frac{1}{\left( 2n+1\right) ^{2}\pi ^{2}}-\frac{1}{\pi ^{2}}%
			\right] ,  \notag
		\end{align}%
	\end{small}
	thus the third term is
	\begin{align}
		& \frac{2}{x}\int_{0}^{x}[\sin \frac{n\pi y}{x}\sin \frac{(n+1)\pi y}{x}%
		y]dy\cos \int_{0}^{t}\frac{E_{n}-E_{n+1}}{\hbar }dt  \notag \\
		& =-\frac{x}{\pi ^{2}}\frac{8n\left( n+1\right) }{(2n+1)^{2}}\cos
		\int_{0}^{t}\frac{E_{n}-E_{n+1}}{\hbar }dt.  \tag{S12}
	\end{align}%
	Then the average position of the small-ball Eq.(\ref{a8}) can be written as
	\begin{equation*}
		\overline{|y|}=\frac{x}{2}-\frac{x}{\pi ^{2}}\frac{8n\left( n+1\right) }{%
			(2n+1)^{2}}\cos \int_{0}^{t}\frac{E_{n+1}-E_{n}}{\hbar }dt, \tag{S13}
	\end{equation*}
	\section*{The Schr\"odinger equation of the fully quantum model in polar coordinates}
	
	The stationary Schr\"odinger equation of the fully quantum model can be
	written as
	\begin{equation}
		\left[ -\frac{\hbar ^{2}}{2M}\frac{\partial ^{2}}{\partial x^{2}}-\frac{%
			\hbar ^{2}}{2m}\frac{\partial ^{2}}{\partial y^{2}}+V(x,y)\right] \Psi
		(x,y)=E\Psi (x,y),  \tag{S14}
	\end{equation}%
	where%
	\begin{equation}
		V(x,y)=\left\{
		\begin{array}{c}
			0,\ \ \ \ 0\leqslant y\leqslant x, \\
			+\infty ,\ \mathrm{otherwise}.%
		\end{array}%
		\right.  \tag{S15} \label{a14}
	\end{equation}%
	Introducing the polar coordinates%
	\begin{equation}
		\sqrt{M}x=\rho \cos \theta ,\sqrt{m}y=\rho \sin \theta ,  \tag{S16} \label{a15}
	\end{equation}%
	and substitution Eq.(\ref{a15}) into Eq.(\ref{a14}) gives%
	\begin{equation*}
		V=\left\{
		\begin{array}{c}
			0,\ \ \ \ 0\leqslant \theta \leqslant \beta =\arctan \sqrt{\frac{m}{M}}, \\
			+\infty ,\ \mathrm{otherwise}.  \tag{S17}
		\end{array}%
		\right.
	\end{equation*}
	Then, the stationary Schr\"odinger equation can be written as
	\begin{equation*}
		\left[ -\frac{\hbar ^{2}}{2M}\frac{\partial ^{2}}{\partial \left( \frac{\rho
				\cos \theta }{\sqrt{M}}\right) ^{2}}-\frac{\hbar ^{2}}{2m}\frac{\partial ^{2}%
		}{\partial \left( \frac{\rho \sin \theta }{\sqrt{m}}\right) ^{2}}+V(\theta )%
		\right] \Psi (\rho ,\theta )=E\Psi (\rho ,\theta ), \tag{S18}
	\end{equation*}%
	and
	\begin{equation*}
		\left[ -\frac{\hbar ^{2}}{2}\left( \frac{\partial ^{2}}{\partial \left( \rho
			\sin \theta \right) ^{2}}+\frac{\partial ^{2}}{\partial \left( \rho \cos
			\theta \right) ^{2}}\right) +V(\theta )\right] \Psi (\rho ,\theta )=E\Psi (\rho ,\theta ). \tag{S19}
	\end{equation*}
	It is easy to find that the above formula is of the form of a two-dimensional
	Schrodinger equation. The Laplace operator in polar coordinates is given by
	\begin{equation}
		\nabla ^{2}=\frac{\partial ^{2}}{\partial \left( \rho \sin \theta \right)
			^{2}}+\frac{\partial ^{2}}{\partial \left( \rho \cos \theta \right) ^{2}}=\frac{\partial ^{2}}{\partial \rho ^{2}}+\frac{1}{\rho }\frac{%
			\partial }{\partial \rho }+\frac{1}{\rho ^{2}}\frac{\partial ^{2}}{\partial
			\theta ^{2}},  \tag{S20}
	\end{equation}%
	The Schr\"odinger equation turns to%
	\begin{equation}
		\left[ -\frac{\hbar ^{2}}{2}\left( \frac{\partial ^{2}}{\partial \rho ^{2}}+%
		\frac{1}{\rho }\frac{\partial }{\partial \rho }+\frac{1}{\rho ^{2}}\frac{%
			\partial ^{2}}{\partial \theta ^{2}}\right) +V(\theta )\right] \Psi (\rho ,\theta )=E\Psi (\rho ,\theta ).  \tag{S21}
	\end{equation}%
	where
	\begin{equation}
		V(\theta )=\left\{
		\begin{array}{c}
			0,\ 0\leqslant \theta \leqslant \beta , \\
			\infty ,\ \mathrm{otherwise}.%
		\end{array}%
		\right.  \tag{S22} \label{a23}
	\end{equation}
	
	\section*{The scattering wave function of the fully quantum model}
	
	The stationary Schr\"odinger equation of the fully quantum model can be
	written as
	\begin{equation}
		-\frac{\hbar ^{2}}{2}(\frac{\partial ^{2}}{\partial \rho ^{2}}+\frac{1}{\rho
		}\frac{\partial }{\partial \rho }+\frac{1}{\rho ^{2}}\frac{\partial ^{2}}{%
			\partial \theta ^{2}})\Psi (\rho ,\theta )+V(\theta )\Psi (\rho ,\theta
		)=E\Psi (\rho ,\theta ).  \tag{S23}
	\end{equation}%
	Substitution the trial solution $\Psi (\rho ,\theta )=R(\rho )\Phi (\theta )$
	into the above formula gives%
	\begin{align}
		-\frac{\hbar ^{2}}{2}\frac{1}{\rho ^{2}}\frac{\partial ^{2}}{\partial \theta
			^{2}}\Phi (\theta )+V(\theta )\Phi (\theta )& =E_{\theta }\Phi (\theta ),
		\tag{S24}  \label{a25} \\
		-\frac{\hbar ^{2}}{2}(\frac{\partial ^{2}}{\partial \rho ^{2}}+\frac{1}{\rho
		}\frac{\partial }{\partial \rho }+\frac{2E_{\theta }}{\hbar ^{2}})R(\rho )&
		=ER(\rho ).  \tag{S24}  \label{a26}
	\end{align}
	Let%
	\begin{equation}
		l^{2}=\frac{2\rho ^{2}E_{\theta }}{\hbar ^{2}},  \tag{S25} \label{a28}
	\end{equation}
	Eq.(\ref{a25}) becomes
	\begin{equation}
		\frac{\partial ^{2}}{\partial \theta ^{2}}\Phi (\theta )+l^{2}\Phi (\theta
		)=0, \ \  (0\leqslant \theta \leqslant \beta ).  \tag{S26}
	\end{equation}
	The solution of the above formula is%
	\begin{equation}
		\Phi (\theta )=\sin l\theta ,  \tag{S27}
	\end{equation}%
	where
	\begin{equation*}
		l=\frac{n\pi }{\beta }, \ \ n=1,2,3\ldots . \tag{S28}
	\end{equation*}
	It follows from Eq.(\ref{a28}) that
	\begin{equation*}
		E_{\theta }=\frac{n^{2}\pi ^{2}\hbar ^{2}}{2\rho ^{2}\beta ^{2}} .\label{a30} \tag{S29}
	\end{equation*}
	Substitution Eq.(\ref{a30}) into Eq.(\ref{a26}) gives%
	\begin{align}
		\frac{\partial ^{2}}{\partial \rho ^{2}}R(\rho )+\frac{1}{\rho }\frac{%
			\partial }{\partial \rho }R(\rho )+(\frac{2E}{\hbar ^{2}}-\frac{1}{%
			\rho ^{2}}\frac{n^{2}\pi ^{2}}{\beta ^{2}})R(\rho ) &=0.  \tag{S30} \label{a35}
	\end{align}%
	Let%
	\begin{equation}
		k^{2}=\frac{2E}{\hbar ^{2}},\xi =k\rho ,  \tag{S31}
	\end{equation}%
	Eq.(\ref{a35}) becomes
	\begin{equation}
		\frac{\partial ^{2}R}{\partial \xi ^{2}}+\frac{1}{\rho }\frac{\partial R}{%
			\partial \xi }+(1-\frac{1}{\xi ^{2}}\frac{n^{2}\pi ^{2}}{\beta ^{2}})R=0.
		\tag{S32}
	\end{equation}%
	As the second kind of Bessel function (Neumann function) tends to $-\infty $ when $\rho
	\rightarrow 0$, the physical solution to the above formula is a linear combination of the first kind of the
	Bessel function, \cite{7}
	\begin{equation}
		R_{l}=C_{l}J_{l}(k\rho ).  \tag{S33}
	\end{equation}%
	For the following calculation, we replace the Bessel function with the
	Hankel function%
	\begin{equation}
		R_{l}=C_{l}\left[ H_{l}^{(1)}(k\rho )+H_{l}^{(2)}(k\rho )\right] ,
		\tag{S34}
	\end{equation}%
	where \cite{7}
	\begin{align}
		H_{\nu }^{(1)}(x) &=\frac{1}{\pi i}\int_{-\infty }^{\infty +\pi i}e^{z\sinh
			t-\nu t}dt,  \tag{S35} \\
		H_{\nu }^{(2)}(x) &=-\frac{1}{\pi i}\int_{-\infty }^{\infty -\pi
			i}e^{z\sinh t-\nu t}dt.  \notag
	\end{align}%
	The no-normalized wave function of this
	system can be written as%
	\begin{equation}
		\Psi_{k,l} (\rho ,\theta )=R(\rho )_{l}\Phi (\theta )=[H_{l}^{(1)}(k\rho
		)+H_{l}^{(2)}(k\rho )]\sin l\theta .  \tag{S36} \label{a41}
	\end{equation}
	
	\section*{The probability current density of the scattering wave function Eq.(18) can be
		divided into two equal and opposite parts without cross term}
	
	The incident and outgoing waves can be written as%
	\begin{align}
		\Psi ^{in}& =H_{l}^{(1)}(k\rho )\sin l\theta ,  \tag{S37} \\
		\Psi ^{sc}& =H_{l}^{(2)}(k\rho )\sin l\theta .  \notag
	\end{align}%
	The wave function is%
	\begin{equation}
		\Psi =\Psi ^{in}+\Psi ^{sc}=\left[ H_{l}^{(1)}(k\rho )+H_{l}^{(2)}(k\rho )%
		\right] \sin l\theta .  \tag{S38}
	\end{equation}%
	The probability current density is
	\begin{equation}
		J=\frac{1}{2}(\Psi ^{\ast }\rho \Psi -\Psi \rho \Psi ^{\ast })=-\frac{i\hbar
		}{2}(\Psi ^{\ast }\nabla \Psi -\Psi \nabla \Psi ^{\ast }),  \tag{S39}
	\end{equation}%
	where%
	\begin{equation}
		\nabla =\rho \frac{\partial }{\partial \rho }\vec{e}_{\rho }+%
		\frac{1}{\rho }\frac{\partial }{\partial \theta }\vec{e}_{\theta} .
		\tag{S40}
	\end{equation}%
	Utilizing the properties \cite{7}
	\begin{align}
		H_{\nu }^{(1)\ast }(x)& =H_{\nu }^{(2)}(x),  \tag{S41} \\
		\frac{\partial }{\partial x}H_{\nu }(x)& =\frac{1}{2}\left[ H_{\nu
			-1}(x)-H_{\nu +1}(x)\right] ,  \notag
	\end{align}%
	one can calculate the probability current density of the incident wave as%
	\begin{equation}
		J^{in}=-\frac{i\hbar }{2}\left( \Psi ^{in\ast }\nabla \Psi ^{in}-\Psi
		^{in}\nabla \Psi ^{in\ast }\right) .  \tag{S42}
	\end{equation}%
	The first term of the above formula is%
	\begin{align}
		\Psi ^{in\ast }\nabla \Psi ^{in}& =\rho \sin ^{2}\left( l\theta \right)
		H_{l}^{(2)}(k\rho )\frac{\partial }{\partial \rho }H_{l}^{(1)}(k\rho )%
		\vec{e}_{\rho }  \notag \\
		& +H_{l}^{(1)}(k\rho )H_{l}^{(2)}(k\rho )\frac{l\sin l\theta \cos l\theta }{%
			\rho }\vec{e}_{\theta }  \notag \\
		& =\frac{k\rho }{2}\sin ^{2}\left( l\theta \right) H_{l}^{(2)}(k\rho )\left(
		H_{l-1}^{(1)}(k\rho )-H_{l+1}^{(1)}(k\rho )\right) \vec{e}_{\rho }
		\notag \\
		& +H_{l}^{(1)}(k\rho )H_{l}^{(2)}(k\rho )\frac{l\sin l\theta \cos l\theta }{%
			\rho }\vec{e}_{\theta }\text{,}  \tag{S43}
	\end{align}%
	and the second one is
	\begin{align}
		\Psi ^{in}\nabla \Psi ^{in\ast }& =\rho \sin ^{2}\left( l\theta \right)
		H_{l}^{(1)}(k\rho )\frac{\partial }{\partial \rho }H_{l}^{(2)}(k\rho )%
		\vec{e}_{\rho }  \notag \\
		& +H_{l}^{(1)}(k\rho )H_{l}^{(2)}(k\rho )\frac{l\sin l\theta \cos l\theta }{%
			\rho }\vec{e}_{\theta }  \notag \\
		& =\frac{k\rho }{2}\sin ^{2}\left( l\theta \right) H_{l}^{(1)}(k\rho )\left(
		H_{l-1}^{(2)}(k\rho )-H_{l+1}^{(2)}(k\rho )\right) \vec{e}_{\rho }
		\notag \\
		& +H_{l}^{(1)}(k\rho )H_{l}^{(2)}(k\rho )\frac{l\sin l\theta \cos l\theta }{%
			\rho }\vec{e}_{\theta }.  \tag{S44}  \label{a48}
	\end{align}%
	Then, Eq.(\ref{a48}) becomes
	\begin{small}
		\begin{align}
			J^{in}& =-\frac{i\hbar }{2}\left( \Psi ^{in\ast }\nabla \Psi ^{in}-\Psi
			^{in}\nabla \Psi ^{in\ast }\right)   \tag{S45} \\
			& =-\frac{i\hbar k\rho }{4}\sin ^{2}\left( l\theta \right) \left[
			H_{l}^{(2)}(k\rho )\left( H_{l-1}^{(1)}(k\rho )-H_{l+1}^{(1)}(k\rho )\right)
			-H_{l}^{(1)}(k\rho )\left( H_{l-1}^{(2)}(k\rho )-H_{l+1}^{(2)}(k\rho
			)\right) \right] \vec{e}_{\rho }.  \notag
		\end{align}
	\end{small}
	Similarly, the probability current density of the radio wave can be
	obtained{\small
		\begin{align}
			J^{sc}& =-\frac{i\hbar }{2}\left( \Psi ^{sc\ast }\nabla \Psi ^{sc}-\Psi
			^{sc}\nabla \Psi ^{sc\ast }\right)   \tag{S46} \\
			& =-\frac{i\hbar k\rho }{4}\sin ^{2}\left( l\theta \right) \left[
			H_{l}^{(1)}(k\rho )\left( H_{l-1}^{(2)}(k\rho )-H_{l+1}^{(2)}(k\rho )\right)
			-H_{l}^{(2)}(k\rho )\left( H_{l-1}^{(1)}(k\rho )-H_{l+1}^{(1)}(k\rho
			)\right) \right] \vec{e}_{\rho }.  \notag
		\end{align}%
	} Then we calculate the probability current density of the wave function.
	Its first item is {\small
		\begin{align}
			\Psi ^{\ast }\nabla \Psi & =\rho \sin ^{2}\left( l\theta \right) \left[
			H_{l}^{(1)}(k\rho )+H_{l}^{(2)}(k\rho )\right] \frac{\partial }{\partial
				\rho }\left[ H_{l}^{(1)}(k\rho )+H_{l}^{(2)}(k\rho )\right] \vec{e%
			}_{\rho }  \notag \\
			& +\left[ H_{l}^{(1)}(k\rho )+H_{l}^{(2)}(k\rho )\right] \left[
			H_{l}^{(1)}(k\rho )+H_{l}^{(2)}(k\rho )\right] \frac{1}{\rho }\sin l\theta
			\frac{\partial }{\partial \theta }\sin l\theta \vec{e}_{\theta }
			\notag \\
			& =\frac{k\rho }{2}\sin ^{2}\left( l\theta \right) \left[ H_{l}^{(1)}(k\rho
			)+H_{l}^{(2)}(k\rho )\right] \left[ H_{l-1}^{(1)}(k\rho
			)-H_{l+1}^{(1)}(k\rho )+H_{l-1}^{(2)}(k\rho )-H_{l+1}^{(2)}\left( k\rho
			\right) \right] \vec{e}_{\rho }  \notag \\
			& +\left[ H_{l}^{(1)}(k\rho )+H_{l}^{(2)}(k\rho )\right] \left[
			H_{l}^{(1)}(k\rho )+H_{l}^{(2)}(k\rho )\right] \frac{l\sin l\theta \cos
				l\theta }{\rho }\vec{e}_{\theta },  \tag{S47}
		\end{align}%
	} the second is {\small
		\begin{align}
			\Psi \nabla \Psi ^{\ast }& =\frac{k\rho }{2}\sin ^{2}\left( l\theta \right) %
			\left[ H_{l}^{(2)}(k\rho )+H_{l}^{(1)}(k\rho )\right] \left[
			H_{l-1}^{(2)}(k\rho )-H_{l+1}^{(2)}(k\rho )+H_{l-1}^{(1)}(k\rho
			)-H_{l+1}^{(1)}\left( k\rho \right) \right] \vec{e}_{\rho }
			\notag \\
			& +\left[ H_{l}^{(2)}(k\rho )+H_{l}^{(1)}(k\rho )\right] \left[
			H_{l}^{(2)}(k\rho )+H_{l}^{(1)}(k\rho )\right] \frac{l\sin l\theta \cos
				l\theta }{\rho }\vec{e}_{\theta }.  \tag{S48}
		\end{align}%
	} So, the probability current density of the wave function is {\small
		\begin{align}
			J& =-\frac{i\hbar }{2}(\Psi ^{\ast }\nabla \Psi -\Psi \nabla \Psi ^{\ast })
			\notag \\
			& =-\frac{i\hbar k\rho }{4}\sin ^{2}\left( l\theta \right) \left(
			H_{l}^{(2)}(k\rho )\left[ H_{l-1}^{(1)}(k\rho )-H_{l+1}^{(1)}(k\rho )\right]
			-H_{l}^{(1)}(k\rho )\left[ H_{l-1}^{(2)}(k\rho )-H_{l+1}^{(2)}(k\rho )\right]
			\right) \vec{e}_{\rho }  \notag \\
			& -\frac{i\hbar k\rho }{4}\sin ^{2}\left( l\theta \right) \left(
			H_{l}^{(1)}(k\rho )\left[ H_{l-1}^{(2)}(k\rho )-H_{l+1}^{(2)}\left( k\rho
			\right) \right] -H_{l}^{(2)}(k\rho )\left[ H_{l-1}^{(1)}(k\rho
			)-H_{l+1}^{(1)}\left( k\rho \right) \right] \right) \vec{e}_{\rho
			}  \notag \\
			& =J^{in}+J^{sc}=0.  \tag{S49}  \label{a54}
		\end{align}%
	} It is clear that, the ecurrent density of the stationary wave function Eq.(18) in the main text can be divided into the incident and outgoing parts, which are opposite, without cross term.

\end{widetext}
\end{appendix}


\end{document}